\documentclass[aps,prd,twocolumn,showpacs,floatfix,superscriptaddress]{revtex4-2}
\usepackage{graphicx}
\usepackage{dsfont}
\usepackage{xcolor}
\usepackage{multirow}
\usepackage[normalem]{ulem}
\usepackage{enumitem} 
\usepackage{lineno}
\usepackage{bm}
\definecolor{refcol}{RGB}{34,34,178}
\usepackage[colorlinks,linkcolor=blue,citecolor=blue,urlcolor=blue]{hyperref}
\usepackage{microtype}
\usepackage{float}
\usepackage[caption = false]{subfig}
\usepackage{appendix}
\usepackage{multirow}
\usepackage{soul}
\usepackage{appendix}
\usepackage{lineno}
\usepackage{amsmath}
\graphicspath{{./Fig/}}

\newcommand{\RNum}[1]{\uppercase\expandafter{\romannumeral #1\relax}}

\begin{document}

\title{Effects of strangeness on the chiral pseudocritical line}

\date{ \today }

\author{Mahammad Sabir Ali}
\email{sabir@niser.ac.in}
\affiliation{School of Physical Sciences, National Institute of Science Education and Research, An OCC of Homi Bhabha National Institute, Jatni-752050, India}

\author{Deeptak Biswas}
\email{deeptakb@niser.ac.in}
\affiliation{School of Physical Sciences, National Institute of Science Education and Research, An OCC of Homi Bhabha National Institute, Jatni-752050, India}

\author{Amaresh Jaiswal}
\email{a.jaiswal@niser.ac.in}
\affiliation{School of Physical Sciences, National Institute of Science Education and Research, An OCC of Homi Bhabha National Institute, Jatni-752050, India}
\affiliation{Institute of Theoretical Physics, Jagiellonian University, ul. St. \L ojasiewicza 11, 30-348 Krakow, Poland}

\author{Hiranmaya Mishra}
\email{ hiranmaya@niser.ac.in}
\affiliation{School of Physical Sciences, National Institute of Science Education and Research, An OCC of Homi Bhabha National Institute, Jatni-752050, India}

\begin{abstract}  
Within a 2+1 flavor Nambu\textendash Jona-Lasinio model, we calculate the curvature coefficients and check them against available lattice QCD estimations. With the observation that the flavor mixing due to the `t Hooft determinant term significantly affects the $\kappa_{2}^{S}$, we explore the effect of $\mu_{S}$ on the $T-\mu_{B}$ crossover lines. With the novel determination of negative $\kappa_{2}^{B}$ at large $\mu_{S}$, we advocate the importance of studying the same in lattice QCD.
\end{abstract}

\maketitle
%


\section{Introduction}
\label{sec:introduction}
The phase diagram of the strongly interacting matter necessitates the determination of the chiral transition line in the high-density and high-temperature regions. The chiral symmetry is broken in the low-density (-temperature) phase of quantum chromodynamics (QCD), which gets restored as the temperature and/or density increases. At vanishing baryon density, the restoration of the chiral symmetry is determined to be a crossover with a pseudocritical temperature $T_{pc} = 156.5 \pm 1.5\ \text{MeV}$ \cite{HotQCD:2018pds}. On the other hand, the transition is expected to be a first order at high density, which is connected to the crossover line through a critical end point (CEP). Although the determination of the crossover line for small values of the baryon chemical potential ($\mu_B$) is quite settled with the recent advancements of lattice QCD (LQCD) \cite{Bellwied:2015rza, HotQCD:2018pds} calculations, the extension of the line at finite $\mu_B$ suffers from the infamous sign problem which leads to the oscillatory behavior of the Monte Carlo sampling method. 


For small chemical potential ($\mu_X$), the pseudocritical line can be Taylor expanded at the lowest order in $\mu_X ^2$, where one defines the line with the following ansatz~\cite{Bellwied:2015rza, Bonati:2018nut, HotQCD:2018pds}:
\begin{equation}
\label{eq:curvature}
 \frac{T_{pc}(\mu_X)}{T_{pc}(0)}=1-\kappa_{2}^{X} \left(\frac{\mu_X}{T_{pc} (0)}\right)^2 - 
 \kappa_{4}^{X} \left(\frac{\mu_X}{T_{pc} (0)}\right)^4~.
\end{equation}
Here, $\mu_X$ corresponds to chemical potential associated with various charges like baryon charge $B$, electric charge $Q$, and strangeness $S$. Such a parametrization allows for the comparison of results from different models and lattice QCD calculations within the same baseline. The curvature coefficients $\kappa_2$ and $\kappa_4$ have been examined by the Taylor expansion method on the lattice~\cite{Gavai:2003mf, Gavai:2004sd, HotQCD:2018pds}. Another standard approach relies on performing the calculations at imaginary chemical potential, followed by an analytic continuation to the real plane~\cite{Bellwied:2015rza, Bonati:2015bha, Borsanyi:2020fev}. The abovementioned results are in good agreement with each other within the respective variances. Similar studies have been performed within the perturbative QCD~\cite{Haque:2020eyj} as well as in the ideal and mean-field hadron resonance gas (HRG) model~\cite{Biswas:2022vat, Biswas:2024xxh} and quark-meson model~\cite{Fu:2019hdw, Schaefer:2004en, Braun:2011iz, Fischer:2012vc, Pawlowski:2014zaa, Fischer:2014ata}. Moreover, the Nambu\textendash Jona-Lasinio (NJL) model has also been employed in this context\cite{Buballa:2003qv} considering two flavors of light quarks.


The effective models, considering the symmetries of the QCD Lagrangian, enable one to probe the matter at extreme conditions like high temperature and/or density, even in the presence of a magnetic field, to understand the phases of the QCD matter and provide a bulk description \cite{Ghosh:2006qh, Pereira:2021xxv}. The NJL model relies on chiral symmetry and provides a qualitative description of the QCD matter considering the pseudoscalar mesons~\cite{Nambu:1961tp, Nambu:1961fr}. Despite the analytical simplicity and the dependence on the parameter sets of such an effective model, the estimations made with the NJL model are quite robust \cite{Hatsuda:1994pi, Rehberg:1995kh}. It acts as a suitable alternative for benchmark estimation at high-density and low-temperature regions \cite{Mishra:2003nr}, as there is no restriction on the applicability of this model at finite density.

Over the past few decades, LQCD and NJL have complemented each other while broadening our understanding of strong interaction in various scenarios. For example, magnetic catalysis (MC) was first shown within an NJL framework~\cite{Klevansky:1989vi, Gusynin:1994re}. Two decades later, lattice QCD not only looked at the MC feature~\cite{Buividovich:2008wf, Buividovich:2009ih, Braguta:2010ej, DElia:2011koc, Bali:2012zg}, but also observed inverse magnetic catalysis around the crossover temperature~\cite{Bali:2012zg}. This results in better versions of NJL models with nonlocal interactions~\cite{Pagura:2016pwr} and external agent-dependent interaction strength~\cite{Farias:2014eca, Ferreira:2014kpa}. Further, in an NJL-like model, the anomalous breaking of $U(1)_A$ symmetry is addressed by explicitly adding the 't Hooft determinant interaction (characterized by coupling $G_d$), which also represents the flavor mixing. Recently, Refs.~\cite{Ali:2020jsy, Ali:2021zsh} explored the effect of $G_{d}$ on isospin-sensitive observables in a two-flavor NJL model and constrained $G_{d}$ using the same from LQCD. In the context of the three-flavor NJL model, $G_{d}$ is the most ill-constrained parameter with a large allowed range while reproducing acceptable values of physical observables~\cite{Hatsuda:1994pi, Rehberg:1995kh}.


In this paper, for the first time in the $2+1$ flavor case with isospin symmetry, the effect of the $G_d$ is explored by incorporating a finite strangeness chemical potential. This provides an opportunity to study the effect of a large $\mu_S$ on the pseudocritical line and provide novel estimations. Although the large variation of $\mu_S$ ($0-200$) MeV is beyond the scope of the freeze-out lines in heavy-ion collisions owing to strangeness neutrality, the present investigation is of particular interest for extending the NJL model at very high density. We organized this paper as follows: In Sec.~\ref{sec:formalism}, we describe the model formalism for a $2+1$ flavor NJL model with the isospin symmetry. We present our results in Sec.~\ref{sec:results} and summarize our findings in Sec.~\ref{sec:summary}.

\section{Formalism}
\label{sec:formalism}


The $2+1$-flavor NJL model Lagrangian is given by
\begin{equation}
  {\cal L}_{\text{NJL}}=\bar{\psi} \left( i\gamma_{\mu}\partial^{\mu}-\hat{m}\right)\psi 
       + {\cal L}_{\text{S}} + {\cal L}_{\text{D}},
  \label{eq:NJL_Lagrangian}
\end{equation}
where the four- and six-point interaction terms are given by 
\begin{equation}
\begin{split}
  {\cal L}_{\text{S}}&=G_{s}\sum_{a=0}^{8}\left[\left(\bar{\psi} \lambda_a\psi\right)^2 + \left(\bar{\psi}\,i \gamma_5 \lambda_a\psi\right)^2\right], \\
  {\cal L}_{\text{D}}&=- G_{d}\left[\det\bar{\psi}_i (1-\gamma_5)\psi_j+\det\bar{\psi}_i (1+\gamma_5)\psi_j\right].
\end{split}
  \label{eq:NJL_Interaction}
\end{equation}
Here, $\psi^{\text{T}}=(u,d,s)$ is the quark triplet in flavor space with an up, down, and strange quark, and $\hat{m}=\text{diag}(m_{u},m_{d},m_{s})$ is the current quark mass matrix. In the interaction, the $\lambda$'s are the Gell-Mann matrices, and in ${\cal L}_{\text{D}}$, the determinant is taken in the flavor space. ${\cal L}_{\text{S}}$ represents the four-quark interaction, with the coupling strength $G_{s}$, which is symmetric under $U(3)\times U(3)$ symmetry. On the other hand, ${\cal L}_{\text{D}}$, with coupling strength $G_{d}$, describes the six-quark interactions known as the 't Hooft determinant. ${\cal L}_{\text{D}}$ is included to break the $U(1)_A$ symmetry explicitly as $U(1)_A$ is anomalous in quantum theory.

To obtain the free energy, it is standard to introduce auxiliary fields using the Hubbard-Stratonovich transformation~\cite{Hubbard:1959ub} to make the Lagrangian quadratic in fermion fields.
%
%
Within mean-field approximation, we can have nonzero vacuum expectation values of these auxiliary fields. In the absence of any other external agents (like a magnetic field, isospin chemical potential, etc.), symmetry only allows the $\bar{\psi}\psi$ channel to acquire nonzero vacuum expectation values, and the mean-field Lagrangian becomes
\begin{equation}
 {\cal L}_{\text{MFA}} = \bar{\psi}\left(i\gamma_{\mu}\partial^{\mu} -\hat{M}\right)\psi - 2G_{s}\sum_{i}\sigma_{i}^2 + 4G_{d}\prod_{i} \sigma_{f},
	 \label{eq:Lagrangian_MFA}
\end{equation}
where $\hat{M}$ is the constituent mass matrix, and the constituent masses are given by~\cite{Hatsuda:1994pi}
\begin{equation}
  M_{i}=m_{i}-4G_{s}\sigma_{i}+2G_{d}\epsilon_{ijk}\sigma_{j}\sigma_{k},
  \label{eq:Constituent_mass}
\end{equation}
with $\sigma_{i}=\langle\bar{\psi}_{i}\psi_{i}\rangle$ being the condensate that works as the order parameter of chiral symmetry breaking. As it is evident from the above equation, $G_{d}$ mixes different flavors.

It is straightforward to integrate out the fermion degrees of freedom from Eq.~\eqref{eq:Lagrangian_MFA} to obtain the free energy. To introduce temperature (T) and chemical potentials ($\mu_{f}$), it is customary to perform the following transformations~\cite{Mustafa:2022got}
\begin{equation}
    p_{0}\rightarrow ip_{4}-\mu_{f}, \qquad p_{4}=(2n+1)\pi T.
\end{equation}
With the above transformation, the integration over $p_{0}$ gets replaced by the sum over Mastubara frequencies, $n$. Moreover, the free energy is given by~\cite{Gastineau:2001zke, Kohyama:2015hix}
\begin{equation}
    \Omega=\Omega_{\text{MF}}+\Omega_{\text{Vac}}+\Omega_{\text{Th}},
    \label{eq:freeenergy}
\end{equation}
where
\begin{eqnarray}
    \Omega_{\text{MF}}&=&2G_{s}\sum_{i}\sigma_{i}^2 - 4G_{d}\prod_{i}\sigma_{i}, \\
    \Omega_{\text{Vac}}&=&-2N_{c}\sum_{i}\int^{\Lambda}\frac{d^3p}{(2\pi)^3}\,\varepsilon_{i}(p), \\
    \Omega_{\text {Th}}&=&-2N_{c}T\sum_{i}\int\frac{d^3p}{(2\pi)^3}\left[ \ln{\left(1+e^{-(\varepsilon_{i}(p)-\mu_{i})/T}\right)}\right.\nonumber\\
    &&+\left.\ln{\left(1+e^{-(\varepsilon_{i}(p)+\mu_{i})/T}\right)}\right].
    \label{eq:omegth}
\end{eqnarray}
With $N_{c}=3$ the number of colors, $\varepsilon_{i}(p)=\sqrt{\Vec{p}^2+M_{i}^2}$ is the energy of the $i$th flavor quark, and $\Lambda$ is the three-momentum cutoff.

To obtain the ground state, one can minimize the free energy defined in Eqs.~\eqref{eq:freeenergy} by solving the following gap equations simultaneously:
\begin{eqnarray}
    \frac{\partial\Omega}{\partial\sigma_u}=\frac{\partial\Omega}{\partial\sigma_d}=\frac{\partial\Omega}{\partial\sigma_s}=0.
\end{eqnarray}
In this study, we have considered the isospin symmetric case; in other words, the electric charge and associated chemical potential ($\mu_{Q}$) are ignored, which implies that $\sigma_{u}=\sigma_{d}= \sigma_{l}$. The quark chemical potential can be written in terms of baryon and strangeness chemical potential
\begin{equation}
\begin{split}
    \mu_{u}&=\mu_{d}=\frac{1}{3}\,\mu_{B} , \\
    \mu_{s}&=\frac{1}{3}\,\mu_{B}-\mu_{S} .
\end{split}
\label{eq:chemicalpotentials}
\end{equation}
Finally, for a fixed $\mu_{B}$ and $\mu_{S}$, we define the pseudocritical temperature ($T_{pc}$) as the inflection temperature where the curvature of $\sigma_l$ changes sign~\cite{Pawlowski:2014zaa}. In the context of LQCD, $T_{pc}$ is generally determined from the maximum of the chiral susceptibility~\cite{Pawlowski:2014zaa}.


\begin{table}[h]
    \centering
    \begin{tabular}{|c|c|c|c|c|c|}
        \hline
        & ~$\Lambda$ (MeV)~ & ~$G_s \Lambda^2$~ & ~$G_d \Lambda^5$~ & ~$m_{l}$(MeV)~ & ~$m_s$(MeV)~ \\
        \hline
         Set I & 631.4 & 1.835 & 9.29 & 5.5 & 135.7 \\
        \hline
         Set II & 602.3 & 1.835 & 12.36 & 5.5 & 140.7 \\
        \hline
    \end{tabular}
    \caption{Parameter sets of the NJL model used in the present work. Set \RNum{1} and \RNum{2} are from Hatsuda and Kunihiro~\cite{Hatsuda:1994pi} and Rehberg {\it et al.}~\cite{Rehberg:1995kh}, respectively.}
    \label{tab:parameter}
\end{table}

Let us note that there are five parameters in this three-flavor NJL model, namely the current quark mass for the strange and light quarks ($m_s$ and $m_l$), two coupling $G_s$ and $G_d$, and the three-momentum cutoff $\Lambda$. After choosing the $m_l = 5.5$ MeV, consistent with chiral perturbation theory~\cite{Gasser:1982ap}, the remaining four parameters are fixed by fitting the pion decay constant and the masses of the pion, kaon, and $\eta'$~\cite{Hatsuda:1994pi,Rehberg:1995kh} to their empirical values. We have considered two widely used parameter sets from Refs.~\cite{Rehberg:1995kh, Hatsuda:1994pi} given in Table~\ref{tab:parameter}. With the parametrization of set \RNum{1}, the mass of the $\eta$ meson is underestimated by $11\%$, while for set \RNum{2}, the same is underestimated by $6\%$. As seen from Table~\ref{tab:parameter}, the dimensionless coupling $G_{d}\Lambda^{5}$ differs by $30\%$ between the two sets, translating into a  $70\%$ variation in $G_{d}$. In this work, we intend to prescribe a way to constrain it more precisely.

\section{Results}
\label{sec:results}
We next consider the thermodynamics of this system to discuss chiral phase transition using Eqs.~\eqref{eq:freeenergy}-\eqref{eq:chemicalpotentials}. For a given value $\mu_{B}$ and $\mu_{S}$, the pseudocritical temperature ($T_{pc}$) is defined to be the inflection point of light quark condensate (the order parameter of chiral symmetry breaking) as a function of the temperature. Before proceeding to investigate the effect of finite $\mu_S$ on the $T-\mu_B$ line, it is essential to check the model estimation against the available lattice QCD results of the curvature coefficients ($\kappa_{2,4}$). Considering $\mu_Q = 0$, we have first investigated the $T-\mu_B$ ($\mu_S=0$) and $T-\mu_S$ ($\mu_B=0$) plane, and find the $\kappa_{2,4}$ by parametrizing the respective pseudocritical lines with the ansatz of Eq.~\eqref{eq:curvature} for the range $\mu_{B,S}/T_{pc}(0)\leq 1.0$ with $T_{pc}(0)=171.1$ and $173.4$ MeV for parameter sets \RNum{1} and \RNum{2}, respectively.

We have tabulated our estimations for the curvature coefficients $\kappa_{2}$ and $\kappa_{4}$ in Tables~\ref{tab:tablek2} and~\ref{tab:tablek4}, respectively. The lattice results are taken from the HotQCD Collaboration~\cite{HotQCD:2018pds} and WB Collaboration~\cite{Bellwied:2015rza}. There is excellent agreement with LQCD estimations for $\kappa_2$ in both the $T-\mu_B$ and $T-\mu_S$ line. We would like to emphasize that the values of $\kappa_2^B$ are similar for both the parameter sets, which infers that the large difference in $G_{d}$ between two parameter sets does not influence the $T-\mu_{B}$ phase line. On the contrary, for the $\mu_B=0$ plane, $\kappa_2^S$ is distinctly different for the two parameter sets. Although these $\kappa_2^S$ values match the lattice estimations within the variances, the estimation with parameter set \RNum{2} has a better agreement with the mean value. The difference in the $\kappa_{2}^{S}$ is attributed to $G_d$, which brings the influence of a strange quark to the light quarks as pointed out in Eq.~\eqref{eq:Constituent_mass}. This motivates us to examine the effect of $G_{d}$ on $\kappa_{2}^{B}$ by exploring the $T-\mu_{B}$ line at various values of $\mu_{S}$.

\begin{table}[h]
    \centering
    \begin{tabular}{|c|c|c|c|}
        \hline
         & ~$\kappa_2^B$ ($\mu_S = 0$)~ & ~$\kappa_2^S$($\mu_B = 0$)~ & $\kappa_2^{B, n_S=0}$ \\
        \hline
        NJL, set \RNum{1} & 0.01627 & 0.01345 & 0.01478 \\
        \hline
        NJL, set \RNum{2} & 0.01619 & 0.01719  & 0.01350 \\
        \hline
        Lattice QCD & 0.016(6) \cite{HotQCD:2018pds}  & 0.017(5)  \cite{HotQCD:2018pds} & 0.012(4)  \cite{HotQCD:2018pds}   \\
                &  &  & 0.0153(18) \cite{Borsanyi:2020fev} \\
        \hline
    \end{tabular}
    \caption{Estimations of $\kappa_2$ for the two parameter sets. The lattice QCD results are taken from Refs.~\cite{HotQCD:2018pds, Borsanyi:2020fev}.}
    \label{tab:tablek2}
\end{table}

\begin{figure}[h!]
\subfloat{\includegraphics[width=0.48\textwidth]{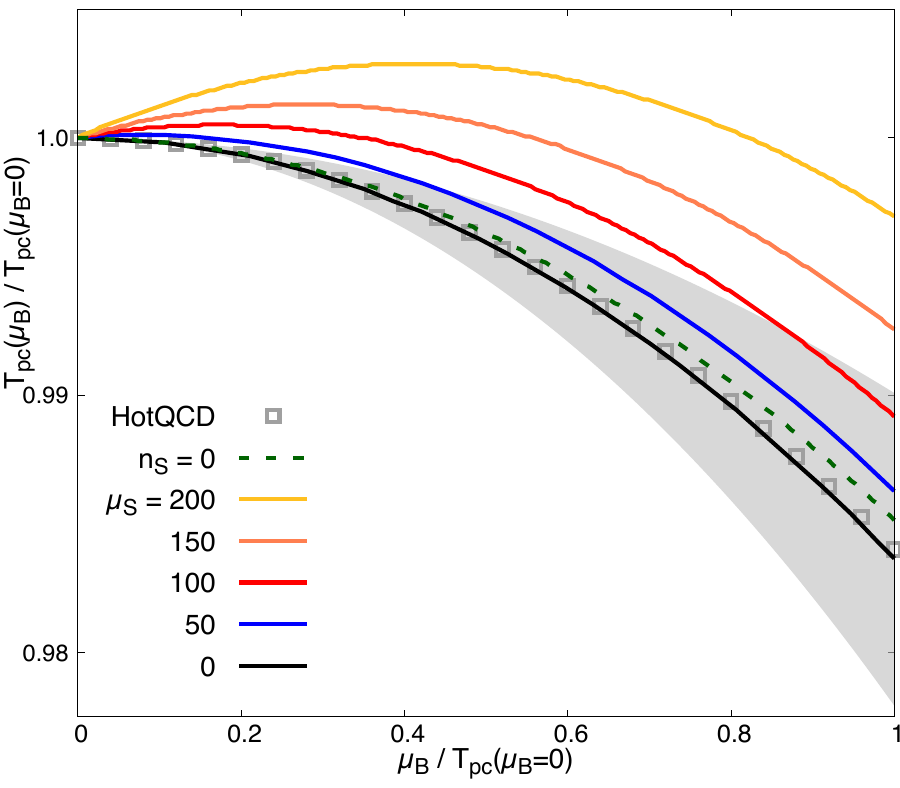}}
\caption{Phase line in the $T-\mu_B$ plane for different $\mu_S$, evaluated with parameter set \RNum{1}. For representative purposes, axes are scaled with respective $T_{pc}(\mu_B=0)$. The continuous lines represent our estimations for various values of $\mu_S$. The data and band are from the HotQCD Collaboration~\cite{HotQCD:2018pds}. The dashed line corresponds to the strangeness neutral case.}
\label{fig:pd}
\end{figure}
To appreciate the effects arising from the strange sector, we have restricted this study to smaller values of baryon chemical potential [up to $\mu_{B}/T_{c}(0)\leq 1.0$] and varied the $\mu_S$ from 0 to 200 MeV. We have restricted the $\mu_S$ within half the kaon mass to exclude the possibility of kaon condensation \cite{Barducci:2004nc}. Because of the difference in the magnitude of the $T_{pc}$ between the NJL and lattice studies, we have scaled the results with their respective $T_{pc}(\mu_{B}=0)$ as shown in Fig.~\ref{fig:pd}. As may be observed, for finite values of $\mu_{S}$, the $T_{pc}$ initially increases with $\mu_B$ and then decreases. For smaller values of $\mu_{B}$, a finite $\mu_S$ decreases the thermal weight in the strange sector [$\mu_S$ comes with a negative sign in the strange thermal distribution; see Eq.~\eqref{eq:chemicalpotentials}] and therefore leads to a higher value of $T_{pc}/ T_{pc}(0)$ for the same $\mu_B$ as shown in Fig.~\ref{fig:pd}. As $\mu_{B}$ increases further, this rise in $T_{pc}$ gets saturated and eventually starts decreasing.

For the first time, such a prominent increase in the pseudocritical temperature ($T_{pc}$) along the $T-\mu_B$ line is observed, which arises due to a finite strangeness chemical potential. This trend was not observed in earlier studies within LQCD~\cite{HotQCD:2018pds, Ding:2024sux} and HRG~\cite{Biswas:2024xxh}, as most of them were performed along the $\mu_S=0$ line or along the freeze-out line, where the strangeness neutrality sets up the limit of $\mu_S \leq \mu_B / 3$~\cite{Ding:2024sux}.

\begin{figure}[h!]
\subfloat{\includegraphics[width=0.5\textwidth]{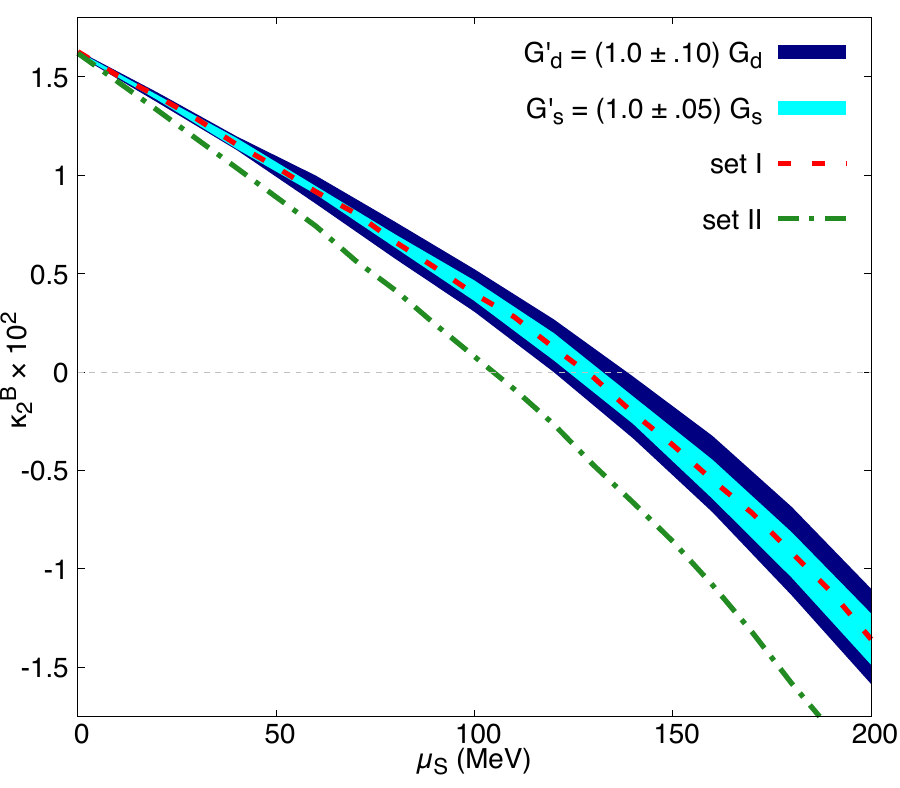}}
\caption{$\kappa_{2}^{B}$ as a function of $\mu_S$ in MeV. The red (dashed) and green (dashed-dot) lines are the central values for sets \RNum{1} and \RNum{2}, respectively. The blue and cyan bands are associated with a $\pm10\%$ change in $G_d$ and a $\pm5\%$ change in $G_s$, respectively, for parameter set \RNum{1}.}
\label{fig:kappa2}
\end{figure}

To quantify the increase in the $T_{pc}$ with $\mu_B$ for a given value of $\mu_S$, we have used the ansatz of Eq.~\eqref{eq:curvature} to extract the curvature coefficients. We have presented the variation of $\kappa_2^B$ with $\mu_S$ in Fig.~\ref{fig:kappa2} for both the parameter sets. The curvature coefficient $\kappa_{2}^{B}$ starts from a positive value for $\mu_S = 0 $ and decreases as we increase the strangeness chemical potential. We wish to emphasize that with $\mu_{S}$, $\kappa_{2}^{B}$ decreases from its positive value at $\mu_{S}=0$ and eventually becomes negative at some $\mu_{S}=\mu_{S}^{c}$. This negative sign of $\kappa_{2}^{B}$ is one of the novel results of the present investigation. This was not observed earlier in the context of the pseudocritical line~\cite{HotQCD:2018pds, Borsanyi:2020fev}. One important observation is that the $\mu_{S}^{\text{c}}$ are distinctively different for the two parameter sets. $G_{s}\Lambda^{2}$ being the same for both the sets, this difference in $\mu_{S}^{c}$ is essentially due to the variance in $G_{d}$. A large $G_{d}$ provides a stronger influence of the strange quark sector on the light quarks, resulting in a faster decrease in $\kappa_{2}^{B}$.

At this juncture, it is instructive to check $\kappa_{2}^{B}$ along the strangeness neutrality line ($n_S = 0$). A finite $\mu_B$ requires the strangeness chemical potential $\mu_S \neq 0$ to achieve zero net strangeness. This corresponds to $\mu_S = \mu_B/3$ as we are considering $\mu_Q = 0$, and there is no vector interaction in the present model. We have found $\kappa_2^{B, n_S=0} < \kappa_2^{B, \mu_S=0} $ as listed in Table.\ref{tab:tablek2}. The decrease of $\kappa_2^{B}$ for the strangeness neutral case is commensurate with the lattice estimations \cite{HotQCD:2018pds, Ding:2024sux} and in accordance with our findings of the reduction of $\kappa_2^{B}$ with $\mu_S$. We would like to comment here that the behavior of the $\kappa_{2}^{B}$ ($\mu_S \neq 0$) is similar to the lattice QCD calculations. The lattice estimations of $n_S= 0$ correspond to values of $\mu_S$, which are not large enough to constrain the flavor mixing determinant coupling. This necessitates LQCD simulations at a larger value of $\mu_S$.


It would be interesting to check the robustness of this negative $\kappa_2^{B}$ on the parametrization of the NJL model itself. For this purpose, we have varied $G_s$, $G_d$ by $5\%$ and $10 \%$, respectively, and examined the effect on the $\kappa_2^{B}$ variation as shown in Fig.~\ref{fig:kappa2}. As discussed earlier, a larger value for $G_d$ increases the coupling between the light and strange sector resulting in a faster decrease of $\kappa_{2}^{B}$. Needless to say, $\kappa_{2}^{B}$ becomes independent of $\mu_{S}$ at $G_{d}=0$ as the strange and light quark sector decouple which is evident in the Lagrangian of the NJL model. On the contrary, the variation of $G_s$ has a weaker effect on the features mentioned above.


\begin{table}[h]
    \centering
    \begin{tabular}{|c|c|c|c|}
        \hline
         & ~$\kappa_4^B$ ($\mu_S = 0$)~ & $\kappa_4^S$ ($\mu_B = 0$) & $\kappa_4^{B, n_S=0}$ \\
        \hline
        NJL, set \RNum{1} & 0.00006 & 0.001477 & 0.000081 \\
        \hline
        NJL, set \RNum{2} & 0.00005 & 0.001892  & 0.000742 \\
        \hline
        Lattice QCD & 0.001(7) \cite{HotQCD:2018pds}  & 0.004(6)  \cite{HotQCD:2018pds} & 0.000(4)  \cite{HotQCD:2018pds}   \\
                &  &  & 0.00032(67) \cite{Borsanyi:2020fev} \\
        \hline
    \end{tabular}
    \caption{Values of $\kappa_4$ for the pseudocritical line for different cases.}
    \label{tab:tablek4}
\end{table}


\begin{figure}[h!]
\subfloat{\includegraphics[width=0.5\textwidth]{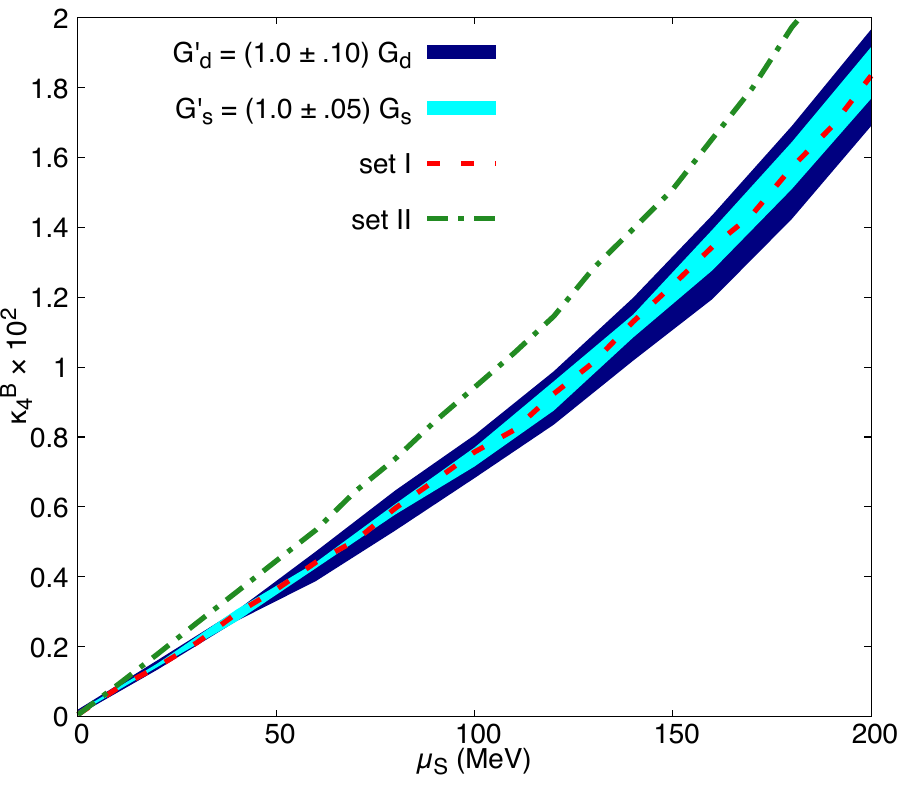}}
\caption{$\kappa_{4}^{B}$ as a function of $\mu_S$ in MeV. Color code is the same as Fig.\ref{fig:kappa2}.}
\label{fig:kappa4}
\end{figure}


Within LQCD, the numerical value of $\kappa_4^{B}$ is consistent with zero \cite{HotQCD:2018pds, Borsanyi:2020fev}, as for the small value of $\mu_{B,S}/T$, the fourth order coefficients of the $\mu_X/T$ expansion are prone to having a weaker effect on the $T- \mu_X$ line. In the present study, we have found $\kappa_4$ to have good agreement for the case (${\mu_B \neq 0,\ \mu_S=0}$) and (${\mu_B=0,\ \mu_S \neq 0}$) as shown in Table~\ref{tab:tablek4}. It would be essential to investigate the same for the $T-\mu_B$ line at various $\mu_S$. For larger values of $\mu_S$, we have found the $\kappa_4$ to be finite (as in Fig.\ref{fig:kappa4}), even with the different parameter sets, as mentioned earlier. These findings suggest that even within the small $\mu_B$ range, a nonzero $\kappa_4$ is possible by switching on a finite strangeness chemical potential $\mu_S$, which is relevant in the context of lattice simulations. 

\section{Summary and conclusion}\label{sec:summary}
In this paper, we have explored the chiral phase boundary of the QCD matter within a $2+1$ flavor Nambu\textendash Jona-Lasinio model with special emphasis on the effect of strangeness on the curvature coefficients $\kappa_2^{B}$ and $\kappa_4^{B}$. To our knowledge, this is the first such exploration within a $2+1$ NJL model. We have considered the isospin symmetric case and $\mu_Q=0$. To have better control over the lowest-order coefficients ($\kappa_2^{X}$), we have limited the study within the range $\mu_X/T \leq 1$. As a benchmark, we have first estimated the $\kappa_{2,4}^{X}$ for three separate cases: (1) the $T-\mu_B$ plane ($\mu_S=0$), i.e., $\kappa_{2,4}^{B}$, (2) the $T-\mu_S$ plane ($\mu_B=0$), i.e, $\kappa_{2,4}^{S}$, and (3) along the strangeness neutrality line $\kappa_{2,4}^{B, n_S=0}$. We have used two standard sets of parametrizations of the $2+1$ NJL model that differ significantly regarding the flavor mixing determinant interaction. Although we have an excellent agreement of $\kappa_{2}^{B}$ with the available LQCD finding for both parameter sets, we have observed that $\kappa_{2}^{S}$ has a strong dependence on the flavor mixing and $U(1)_A$ breaking 't Hooft interaction. Between the two parameter sets used, set \RNum{2} with a higher value of $G_{d}$ reproduces the lattice estimation of $\kappa_{2}^{S}$ better. To explore the effects of flavor mixing through $G_{d}$, it is instructive to study the $\mu_S$ dependence of the $T-\mu_B$ lines, which have been quantified by estimating $\kappa_2^{B}$ as a function of $\mu_S$.

It is interesting to note that, we have observed for the first time the decreasing behavior of $\kappa_{2}^{B}$ with $\mu_{S}$, which is interesting and interpreting it in the framework of the NJL model is also relevant. More importantly, we have found that it becomes negative for sufficiently large values of $\mu_{S}$. Further, it is also observed that the value of $\mu_S$ where $\kappa_{2}^{B}$ vanishes is different for the two parameter sets. This difference is attributed to the fact that a larger value of $G_{d}$ strengthens the strange contribution to the light sector, resulting in a faster decrease. We expect that the outcomes from LQCD investigations for $\kappa_{2}^{B}$ at large enough $\mu_S$ will assist in better constraining the 't Hooft coupling $G_{d}$, thereby enhancing our understanding of effective models like NJL and the underlying QCD. In this article, we have prescribed a way to quantify the flavor mixing, which is an important development toward understanding the effective model and, eventually, QCD.


At this juncture, we note that the physical scenarios accessible in the present heavy-ion collision experiments are rather constrained to $n_S=0$ and $n_Q = 0.4n_B$. However, from a theoretical perspective, it is possible to explore QCD in all directions as it helps one to calibrate and understand various aspects of the theory. Lattice QCD has explored the phase diagram at finite $\mu_Q$ and $\mu_S$~\cite{HotQCD:2018pds}, and recently, it has been extended toward a larger value of strangeness chemical potential~\cite{Ding:2024sux}. Our study provides an alternate approach in this direction. We have investigated here the low $\mu_B$ region of the phase diagram for the study of curvature coefficients. The effect of flavor mixing on CEP in the presence of finite $\mu_{S}$ will be interesting and deserves a separate investigation which will be explored in a future work.

\section*{Acknowledgments}
D.B. is supported in part by the Department of Science and Technology, Government of
INDIA under the SERB National Post-Doctoral Fellowship Reference No. PDF/2023/001762. M. S. A. and D. B. would like to extend thanks to C. A. Islam for the fruitful discussions and critical reading of the manuscript. 

\bibliography{ref_njl3f}

\end{document}